\documentclass{PoS}

\newcommand{\be}{\begin{equation}}
\newcommand{\ee}{\end{equation}}
\newcommand{\B}{\rm B}

\newcommand{\gBB}{g_{{\rm B^* B} \pi} }
\newcommand{\Nf}{N_{\rm f}}
\newcommand{\Mb}{M_{\rm b}}

\title{Standard Model parameters and heavy quarks on the lattice}

\ShortTitle{SM parameters and heavy quarks on the lattice}

\author{\speaker{Michele Della Morte}\\
        CERN, Physics Department, TH Unit, CH-1211 Geneva 23, Switzerland\\
        E-mail: \email{michele.della.morte@cern.ch}}

\abstract{
\vspace{-9.5cm}
I review recent progresses in heavy quarks physics on the lattice. 
I focus on decay constants and form factors relevant for the extraction of CKM 
matrix elements from experimental data. $B-\bar{B}$ mixing is also discussed. 
In the last part of the paper I describe phenomenological 
applications  of Heavy Quark Effective Theory (HQET) on the lattice, 
presenting in some detail the recent 
non-perturbative determination of the b-quark mass including O$(1/m_{\rm b})$
corrections.
\vspace{-11.5cm}
\begin{flushright}
CERN-PH-TH/2007-205
\end{flushright}
}

\FullConference{The XXV International Symposium on Lattice Field Theory\\
		 July 30-4 August 2007\\
		 Regensburg, Germany}

\begin{document}

\section{Introduction}
Next year CERN's Large Hadron Collider (LHC) will start delivering proton beams
for physics collision. The LHCb experiment is designed to exploit the
enormous LHC potential in the b-quark sector for measurements of the CKM
parameters to such a high precision that possible contributions from
TeV-scale New Physics to the mixing mechanism will become visible. To give
an idea, the production of b hadrons at LHCb is expected with the annual
yield of $10^{12}$ b-$\overline{\rm b}$ pairs~\cite{Barsuk:2005ac}. 
Possible future super-B factories would further extend the set of high
precision b-physics measurements~\cite{Akeroyd:2004mj}.

This programme can provide a stringent test of the Standard Model and 
potentially lead to the discovery of New-Physics only if at the same time a
significant progress on the theory side is made.
To get a flavor about the required precision it is useful to have a look at
the experimental and theoretical situation  for a few low-energy
flavor-violating observables where non-Standard effects were expected to 
contribute.

Let us start with the inclusive radiative B-meson decay. The world average 
performed by the Heavy Flavor Averaging Group~\cite{Barberio:2006bi}
 for $E_{\gamma}>1.6$ GeV yields the branching ratio
\be
{\mathcal{B}}(\bar B \rightarrow X_{\rm s} \gamma)=
(3.55 \pm 0.24^{+0.09}_{-0.10}\pm 0.03) \times 10^{-4} \;,
\ee
to be compared with the Standard Model NNLO analysis of 
Ref.~\cite{Misiak:2006zs}, which for the same cut on the photon energy gives
\be
{\mathcal{B}}(\bar B \rightarrow X_{\rm s} \gamma)=
(3.15 \pm 0.23) \times 10^{-4} \;.
\ee
The values are consistent basically within one (combined) sigma (about 10\%),
which implies that the difference between the Standard Model (SM) and the 
experimental numbers can be of order 20\%. Notice that this estimate
does not depend on theoretical inputs from the lattice.

New Physics in principle can be found also in purely leptonic $B^{\pm}$ 
(or $D^{\pm}$)
decays, which can be enhanced by charged Higgs exchange contributions in 
any model with two Higgs doublets~\cite{Hou:1992sy}. Again, the average of the
experimental numbers for ${\mathcal{B}}(B \to \tau\nu)$ from Belle and 
Babar~\cite{Ikado:2006un,Aubert:2006fk} 
is in good agreement with  the SM theoretical computation although the total 
error is above 30\% in the first case and around 20\% in the second.
On the theory side this is mainly due to the uncertainty on $V_{\rm ub}$
and on the decay constant $F_{\rm B}$, which will be discussed in more detail in
the next section. The experimental error on the other hand is expected to 
decrease in the future. At a Super-B factory with hundred times the luminosity
of Belle the branching ratio would be measured with a precision of 3\%.

The leptonic decays of the neutral $B_{(\rm s)}$ meson are very rare 
(the SM branching ratio is O($10^{-9}$)) and they haven't been observed so far.
The most recent experimental upper bound on 
${\mathcal{B}}(B_{\rm s} \to \mu^+ \mu^-)$ is $1 \times 10^{-7}$ from CDF~\cite{CDF_Bmumu}.
This decay is included among the LHCb physics goals with an SM expectation of 
20 events per year~\cite{Barsuk:2005ac}.
There is quite some excitement around this channel as it can be 
significantly enhanced in various extensions of the Standard Model.
For example in the Minimal Supersymmetric Standard Model with large 
$\tan\beta$ (where $\tan\beta$ is the ratio of the two neutral Higgs field 
vacuum expectation values) the enhancement can be up to three
orders of magnitude compared to the SM. That is due to the appearance of flavor
changing couplings of the neutral Higgs bosons generated by non-holomorphic 
terms after supersymmetry breaking~\cite{Babu:1999hn}.

Finally New Physics might contribute to $D-\bar{D}$ mixing, which has been
 recently observed by Babar~\cite{Aubert:2007wf}. It is hard to quantify the
 possible size of non-Standard effects here as the SM theoretical predictions 
are very uncertain. It is also not clear whether useful quantities can be 
computed on the lattice to describe the process, which is affected by 
long-distance contributions that are not captured by the Operator Product
Expansion (see~\cite{Petrov:2006nc} for a more exhaustive discussion).

From the examples above I conclude that to keep the pace with experiments and
help in the search of New Physics lattice computations must aim at high 
precision, typically between a few percent and 10\% depending on the process
and the corresponding non-perturbative parameters needed.
In order to achieve such an accuracy the computations must start from first
principles, which implies that the light fermions must be treated as dynamical degrees of freedom, and all the 
systematics associated with renormalization,  extrapolation to the continuum limit and
chiral  extrapolation must be kept under control. In the rest of the review 
I will try to show that each of those effects can introduce an uncertainty of
 O(5\%), and I will do that while presenting a selection of recent results for
 $B$ and $D$ meson decay constants, the $B_{\rm B_{\rm(s)}}$ parameter and semi-leptonic 
form factors for heavy-light and heavy-heavy transitions.

In the last part I will describe an approach (not necessarily the only one) in
which all these systematics  can be addressed non-perturbatively.
As an application I will present the (quenched) computation of the b-quark mass
in HQET including O$(1/m_{\rm b})$ effects~\cite{Della Morte:2006cb}.

\section{$B_{\rm(s)}$ and $D_{\rm (s)}$ meson decay constants}
In the Standard Model the purely leptonic decays of charged $B$ and $D$ mesons
proceed via quark annihilation into a $W$ boson.
Taking as example the $B \to \tau \nu_{\tau}$ channel, the branching ratio 
can be parameterized as
\begin{equation}
{\cal{B}}(B^- \rightarrow \tau^-\bar{\nu}_{\tau}) \propto F_{\rm B}^2 |V_{\rm ub}|^2 \;,
\label{decay}
\end{equation}
which turns out to be O$(10^{-4})$. The proportionality factor is a function 
of well-known masses, life-times and the Fermi constant. In eq.~(\ref{decay}) $F_{\rm B}$ is the 
$B$ meson decay constant, which is given by the matrix element of the
 heavy-light axial current between the vacuum and the $B$-meson state, while
$V_{\rm ub}$ is the relevant element (actually the smallest and least known)
 of the CKM matrix. Similarly $F_{\B_{\rm s}}$ is the non-perturbative matrix element
necessary for the SM prediction of the $B_{\rm s} \to \mu^+ \mu^-$ branching ratio discussed
in the Introduction.

\begin{figure}[htb]
\vspace{-.8cm}
\begin{center}
\includegraphics[width=8.65cm,height=9.4cm,angle=-90]{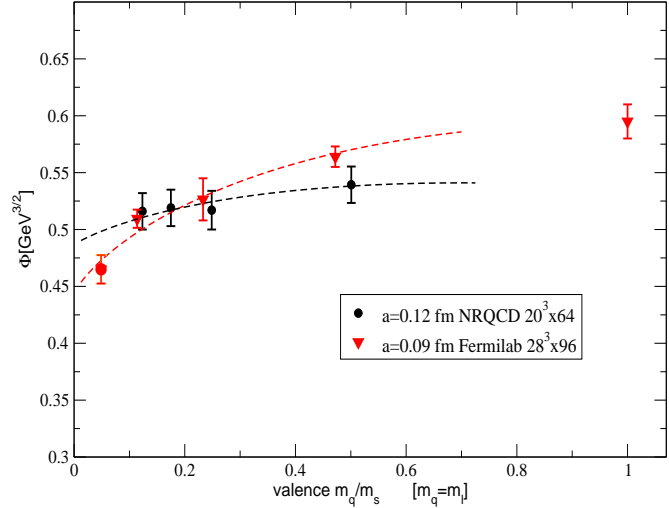}
\caption{S$\chi$PT-guided extrapolations for $F_{\rm B}$ from~\cite{Gray:2005ad,Simone06}.}
\label{Fb}
\end{center}

\vspace{-.5cm}
\end{figure}

The $B$ meson decay constant has been computed with three dynamical flavors
by the HPQCD and the Fermilab, MILC 
Collaborations~\cite{Gray:2005ad,Simone06,Simone07}. In both cases
the rooted staggered quarks configurations generated by the MILC Collaboration
with the AsqTad action have been employed~\cite{Aubin:2004wf}.
The heavy b-quark is simulated by using NRQCD in~\cite{Gray:2005ad}
and the Fermilab action in~\cite{Simone06,Simone07}.
The results from~\cite{Gray:2005ad,Simone06} for 
$\Phi_{\rm q}=F_{\rm B_{\rm q}}\sqrt{m_{\rm B_{\rm q}}}$ 
are shown in figure~\ref{Fb} as a function of the sea quark mass in units
of the strange quark mass  and for the unitary (light sea quark mass equal to the light
 valence quark mass) points only.
The curves are the Staggered Chiral Perturbation Theory 
(S$\chi$PT)~\cite{Aubin:2005aq} fits.
Although the same formulae have been used and the lattice resolutions
are not too coarse, the results suggest quite different chiral behaviors
(reflected in a 5\% difference on the ratio $F_{\rm B_{\rm s}}/F_{\rm B}$), probably due 
to residual cutoff effects. It is interesting to note 
the consistency of the Fermilab data with the curvature predicted from S$\chi$PT, 
notice however  that there the coupling $\gBB$ appearing in the non-analytic terms
has been set to $g_{{\rm D^*D}\pi}$ from the CLEO experiment before performing the fit.

The final result quoted in~\cite{Gray:2005ad} is $F_{\rm B}= 216(9)(19)(4)(6)$ MeV, where
the first error is statistical (including chiral extrapolations) and the others are 
estimates of the systematics. The largest
one in particular is due to the matching  between the heavy-light current in QCD and
in NRQCD. This matching involves power divergent mixings between dimension-three and
dimension-four operators in the effective theory and the subtraction has been performed 
by considering the one-loop contribution only. 
The other systematics included are discretization effects and relativistic corrections.
Most of these cancel in the ratio $F_{\rm B_{\rm s}}/F_{\rm B}$, for which
the value $1.20(3)(1)$ is obtained. 

The Fermilab Collaboration in~\cite{Simone06} preferred to quote numbers for the ratio only
as at that time the computation of the relevant renormalization constants
was not yet completed. The result is $F_{\rm B_{\rm s}}/F_{\rm B}=1.27(2)(6)$
where the second uncertainty is mainly due to the chiral extrapolation.
An update including results from two additional lattice resolutions ($a=0.12$ and $0.15$ fm)
and the use of the  matching renormalization constants computed at one-loop in~\cite{Aida07} 
has been presented at this conference~\cite{Simone07}. The preliminary analysis yields
$F_{\B}= 191(5)(8)$ MeV and $F_{\B_{\rm s}}/F_{\B} = 1.30(3)(4)$, both in good agreement
with the NQRCD results.

\begin{figure}[htb]
\begin{center}
\includegraphics[width=7.8cm,height=7.8cm]{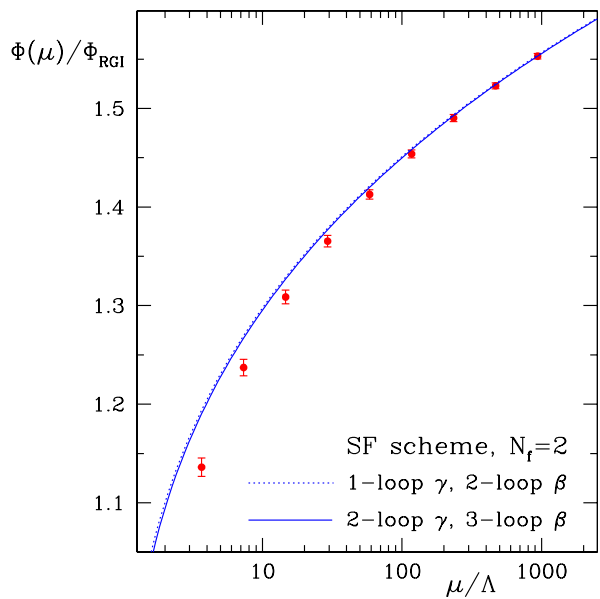}
\caption{Non-perturbatively computed running matrix element of the static-light axial
current in the SF scheme from~\cite{Della Morte:2006sv}. The dotted and solid lines are
obtained from perturbation theory using 1/2 and 2/3 loops expressions for the 
anomalous dimension of the current~\cite{Kurth:2000ki} and the 
$\beta$-function~\cite{Bode:1999sm}.
As an information, the $\Lambda$ parameter from~\cite{DellaMorte:2004bc} is 
$\Lambda_{\rm SF} \simeq 100$ MeV.}
\label{SF_ax_run}
\end{center}

\vspace{-.2cm}
\end{figure}

The ALPHA Collaboration has completed the non-perturbative computation of the 
renormalization constant of the static-light axial current with two dynamical flavors
in the Schr\"odinger functional (SF) scheme~\cite{Della Morte:2006sv}. The main result
is the universal (i.e. regularization independent) factor $\Phi(\mu)/\Phi_{\rm RGI}$
relating a matrix element $\Phi(\mu)$ of the static-light axial current renormalized 
at the scale $\mu$ to its scheme-independent (Renormalization Group Invariant) version.
The result is shown in figure~\ref{SF_ax_run}.
For $\mu \leq 2$ GeV perturbation theory fails in reproducing the correct 
result and there would be no way to detect it within perturbation theory only, as the 
convergence of the series appears to be very good in all the range plotted.
At the most non-perturbative scale, where large volume matrix elements relevant for
phenomenology are usually renormalized, the discrepancy reaches $5\%$. 
The regularization dependent constants needed to match the bare matrix elements
to the ones renormalized at this scale have also been computed in~\cite{Della Morte:2006sv}  
for different static actions
(see~\cite{Della Morte:2005yc} for
their precise definition) and for the range of bare couplings relevant for simulations
in large volume using Wilson-Clover fermions.

As a first application $F_{\B_{\rm s}}^{\rm stat}$ has been computed
on a $24^3 \times 32$ lattice with $a\simeq 0.08$ fm and (degenerate) sea quark masses 
close to the strange quark mass. The result $F_{\B_{\rm s}}^{\rm stat}=297(14)$ MeV is rather 
large compared  for example to the quenched value $F_{\B_{\rm s}}=193(6)$ MeV 
obtained in~\cite{fBstat_Nf0} 
by linearly interpolating in the inverse meson mass between continuum results in the static 
approximation and in the relativistic theory with heavy quarks around the charm. Several effects
may concur in producing the large $N_{\rm f}=2$ number, for instance cutoff effects, 
$1/m_{\rm b}$ corrections or sea quark mass effects. While to estimate the latter it is 
necessary to repeat the computation at lighter sea quark masses, for the first two an impression
can be gathered by comparing with the static result at a similar  lattice spacing in the
quenched approximation, which turns out to be 
$F_{\B_{\rm s}}^{\rm stat}(\Nf=0,a\simeq 0.08$ fm)$= 247(5)$ MeV from~\cite{fBstat_Nf0}.
This still leaves room for sizeable effects of the dynamical fermions, which I will consider 
again in the following when discussing the $D_{\rm (s)}$ meson decay constant. 
Remaining within the quenched approximation  $F_{\B_{\rm s}}$
has also been computed including $1/m_{\rm b}$ corrections explicitly in HQET~\cite{Garron07}.
The final result is nicely consistent with the one obtained by the interpolation discussed
above, although with larger errors. The computation will be described in more detail 
in the last section.

Let us now consider the $D_{(\rm s)}$ system. The decay constants $F_{\rm D}$ and 
$F_{\rm D_{\rm s}}$ can be used to extract the CKM matrix elements $V_{\rm cd}$ 
and $V_{\rm cs}$ from the CLEO data~\cite{Stone:2007dy}. The most recent computation
by the HPQCD Collaboration~\cite{Follana:2007uv} includes the effects of 2+1 dynamical flavors
implemented in the staggered AsqTad formalism by use of the fourth root of the quark determinant.
For the valence fermions two different variants of the new Highly Improved Staggered Quark 
(HISQ) action~\cite{Follana:2006rc} have been used  for the light (including strange) and the 
charm quarks. Some simulations parameters are collected in table~1 while results 
are shown in figure~3, taken from~\cite{Follana:2007uv}.
%
\begin{minipage}{6cm}
\begin{center}
\begin{tabular}{ccc}
\hline
$V$             &    $a$  &   $am_c$ \\
\hline
$16^3\times 48$ & 0.15 fm &   $0.85$ \\
$20^3\times 64$ & 0.12 fm &   $\simeq 0.65$ \\
$24^3\times 64$ & 0.12 fm &   $\simeq 0.65$ \\
$28^3\times 96$ & 0.09 fm &   $\simeq 0.43$ \\
\hline 
\end{tabular}
\setcounter{table}{1}

\vspace{0.15cm}
{\footnotesize {\bf Table 1}: Lattice volumes, lattice \\
spacings $a$ and values of the charm \\
quark mass in units  of $a$
from~\cite{Follana:2007uv}.}
\end{center}
\end{minipage}
\begin{minipage}{9.05cm}
\vspace{0.12cm}
The values of the charm quark mass in units of $a$ are quite large in this study and they
might cause some concern about the size of cutoff effects. In figure~3 however 
these appear to be roughly at the 10\% level at the coarsest lattice resolution. The concern is then
whether all the data are in the scaling region and a continuum limit extrapolation is justified or not.
It would therefore be desirable to repeat the computation at the very fine resolution $a=0.06$ fm where
the MILC Collaboration is $\phantom{,}$indeed producing configurations.
\vspace{0.12cm}
\end{minipage}
The final result $F_{\rm D_{\rm s}}=241(3)$ MeV,
$F_{\rm D_{\rm s}}/F_{\rm D}=1.162(9)$ is obtained by performing a simultaneous chiral and continuum
 extrapolation of the data at different quark masses and lattice spacings. 
\begin{minipage}{7cm}
\vspace{0.12cm}
The overall error includes corrections due to the $u/d$ quark
mass difference and electromagnetic effects (see table~2 in~\cite{Follana:2007uv}  
for the detailed error budget), which make the claimed precision clearly impressive. In my opinion such a 
precision calls for a complete clarification of the issues related to the use of the ``fourth root trick''
in dynamical simulations of staggered quarks.
The discussion on the localization, the unitarity and 
the symmetry content of the ``rooted'' theory~\cite{Bunk:2004br,Sharpe:2006re,Creutz:2007rk,Kronfeld07} 
is still ongoing and a final conclusion in favor or disfavor
of it hasn't been reached yet. Also, to be able to conclusively judge on the error budget
it would be useful to have more details concerning the Bayesian fits performed, the precise functional forms used
in the continuum/chiral extrapolations and also some algorithmic details. Simulations are indeed described
for sea quark masses above one fifth of the strange quark mass only~\cite{Aubin:2004wf}. 
Some of these points will probably be clarified in the longer publication announced in~\cite{Follana:2006rc}.
\end{minipage}
%
\begin{minipage}{8cm}
\begin{center}
\vspace{0.2cm}
\includegraphics[width=7.7cm,height=7.85cm]{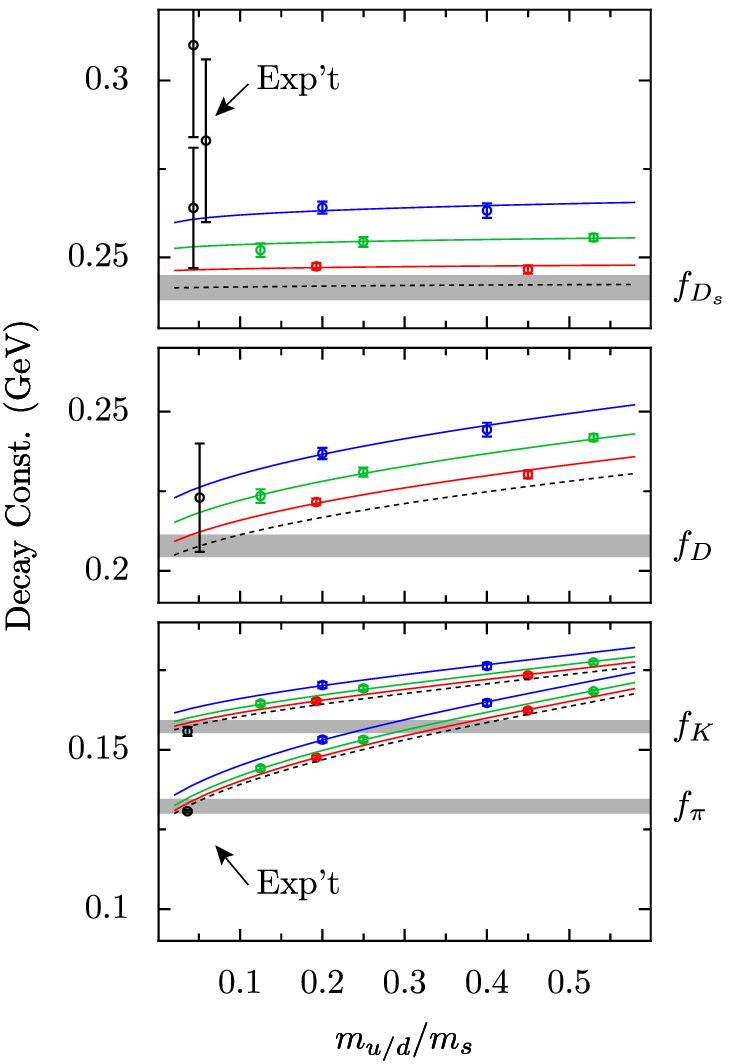}
\footnotesize{{\bf Figure 3:} Results for the $D$, $D_{\rm s}$ ($K$ and $\pi$) decay \\ 
constants from~\cite{Follana:2006rc}
for three lattice resolutions \\ (see table~1). The chiral fits are performed together \\ with those of the 
corresponding meson masses. \\ The continuum limit is given by the dashed lines \\ and the final, chirally
extrapolated, results are \\ represented by the shaded bands.}
\setcounter{figure}{3}
\end{center}
\end{minipage}

\begin{figure}[htb]
\begin{center}
\includegraphics[width=4.9cm,height=6.9cm,angle=-90]{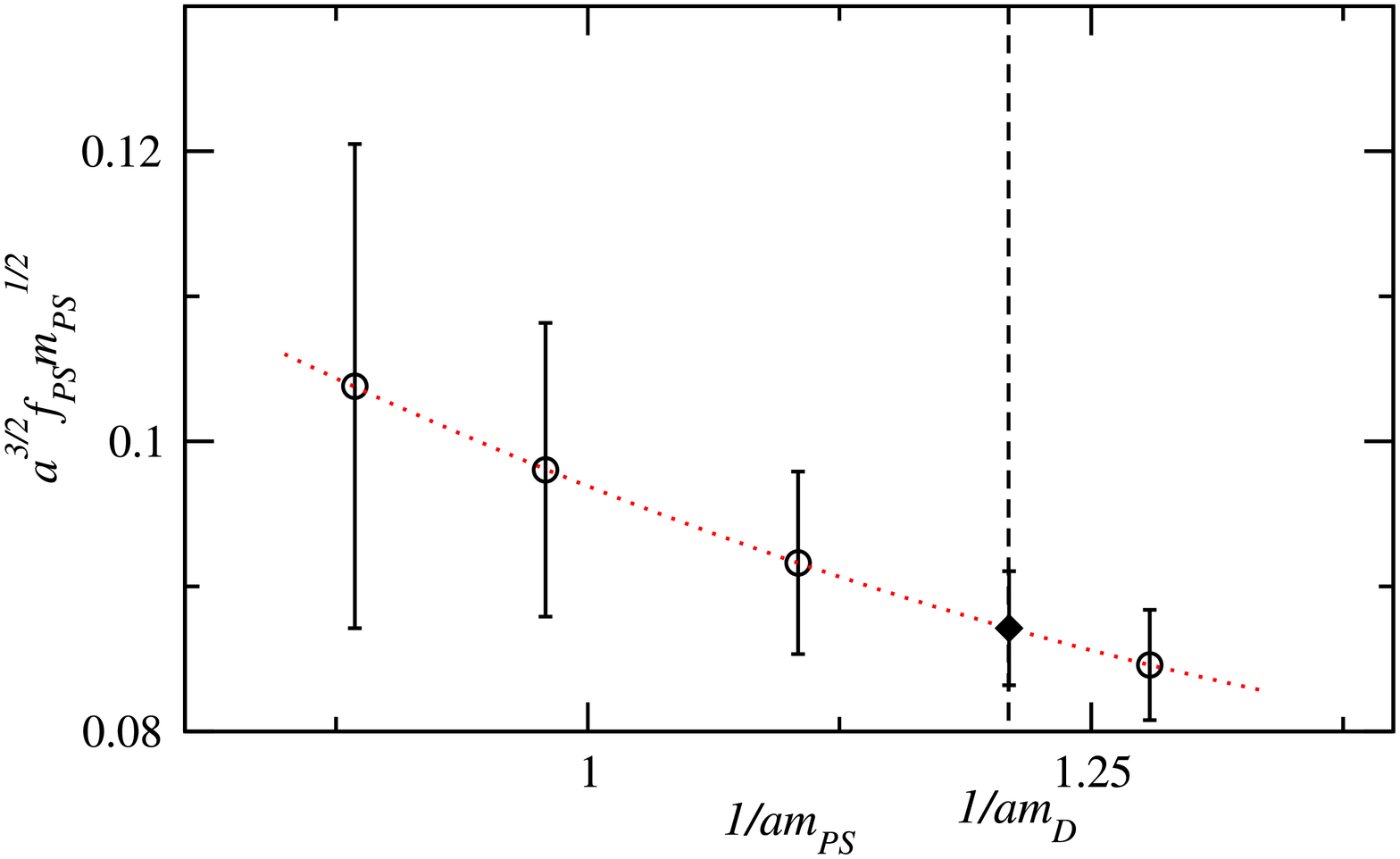}
\caption{Scaling of $F_{\rm PS}\sqrt{m_{\rm PS}}$ as a function of the inverse pseudoscalar meson mass 
$1/m_{\rm PS}$. Plot from~\cite{Blossier:2007pt}.}
\label{fig:tmQCD}
\end{center}

\vspace{-.5cm}
\end{figure}

The European Twisted Mass (ETM) Collaboration has presented at this conference an application in the charm sector
of the twisted mass (tmQCD) formalism with two dynamical light flavors~\cite{Blossier:2007pt}. By working at maximal 
twist the quantities computed are automatically O($a$) improved~\cite{Frezzotti:2003ni} and no renormalization 
constants have to be calculated to obtain the decay constants, as first pointed out in~\cite{Frezzotti:2001du}.
Configurations have been generated for two lattice volumes $24^3 \times 48$ and $32^3 \times 64$ with
lattice spacings $a\simeq 0.09$ and $0.07$ fm respectively. The sea quark masses are in the range $m_{\rm s}/6$
and $2 m_{\rm s}/3$. The decay constants $F_{\rm D}$ and $F_{\rm D_{\rm s}}$ have been obtained by 
interpolating to the proper value of the meson mass the results produced for heavy quarks around the charm.
The interpolation in the case of the $D$ meson at the coarser lattice resolution is shown in
figure~\ref{fig:tmQCD}, taken from~\cite{Blossier:2007pt}. In this case four points have been fitted with 
a three-parameters functional form inspired by HQET.
The preliminary results quoted are $F_{\rm D_{\rm s}}=271(6)(4)(5)$ MeV  and 
$F_{\rm D_{\rm s}}/F_{\rm D}=1.35(4)(1)(7)$ for $a\simeq 0.09$ fm. The second error comes from the uncertainty on
the strange quark mass, while the third is due to the uncertainty on the lattice spacing in the case of
 $F_{\rm D_{\rm s}}$ and to the chiral extrapolation in the case of the ratio 
$F_{\rm D_{\rm s}}/F_{\rm D}$. The determinations at the finer lattice resolution provide 
consistent results though with larger errors. It is important to assess precisely the size of cutoff effects
on the result above, as that is obtained interpolating in pseudoscalar meson masses which are very close
to the cutoff scale (see figure~\ref{fig:tmQCD}).

\begin{figure}[htb]
\begin{center}
\includegraphics[width=6.9cm,height=4.9cm]{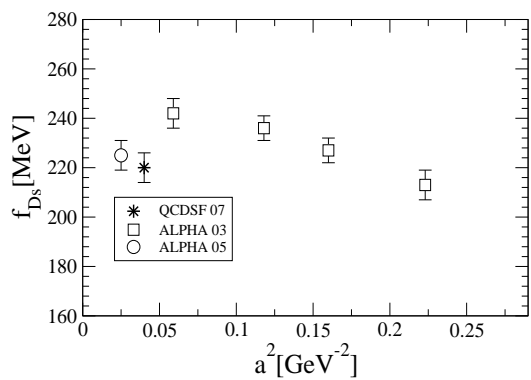}
\caption{The $D_{\rm s}$ meson decay constant versus $a^2$ with non-perturbatively O($a$) improved Wilson 
fermions in quenched QCD. Results from~\cite{Ali Khan:2007tm} (star) and~\cite{Juttner:2003ns,Juttner:2004tb}
(squares and circle).}
\label{fig:QCDSF}
\end{center}

\vspace{-.5cm}
\end{figure}

Finally, in~\cite{Ali Khan:2007tm}, the QCDSF Collaboration calculated the decay constants of heavy-light
pseudoscalar mesons on a very fine lattice ($a\simeq 0.04$ fm, $V=40^3 \times 80$) using
non-perturbatively O($a$) improved Wilson fermions in the quenched approximation.
The result for $F_{\rm D_{\rm s}}$ is presented
in figure~\ref{fig:QCDSF} together with those obtained by the ALPHA collaboration using the same
action but in a larger range of lattice resolutions~\cite{Juttner:2003ns,Juttner:2004tb}.
The agreement between the results is quite satisfactory and suggests the possibility of a joint continuum
extrapolation (excluding for example the point at the coarsest lattice spacing).
The computation of  the $D$ meson decay constant requires a chiral extrapolation, which in~\cite{Ali Khan:2007tm}
has been performed by linearly extrapolating data corresponding to ``pion'' masses above  500 MeV. An 
uncertainty associated with this chiral extrapolation is not estimated for the final error budget
and the values $F_{\rm D_{\rm s}}=220(6)(5)(11)$ MeV (the third error is ascribed to a 10\% ambiguity
in the lattice spacing)  and $F_{\rm D_{\rm s}}/F_{\rm D}=1.068(18)(20)$ are eventually obtained.

The decay constants of B-mesons are also computed in~\cite{Ali Khan:2007tm}. In this case however
bare quark masses $m_{\rm q}$ with $am_{\rm q} \simeq 0.7$ need to be considered and the residual, O($a^2$),
cutoff effects on $F_{\rm B_{\rm s}}$ are estimated by the authors of~\cite{Ali Khan:2007tm} to be 12\%. This 
sets the limits of the approach. In addition, for such masses roundoff effects on the quark propagator at
large time separations should be carefully checked as well~\cite{Juttner:2005ks}.

\begin{figure}[htb]
\begin{center}
\includegraphics[width=9cm,height=6cm]{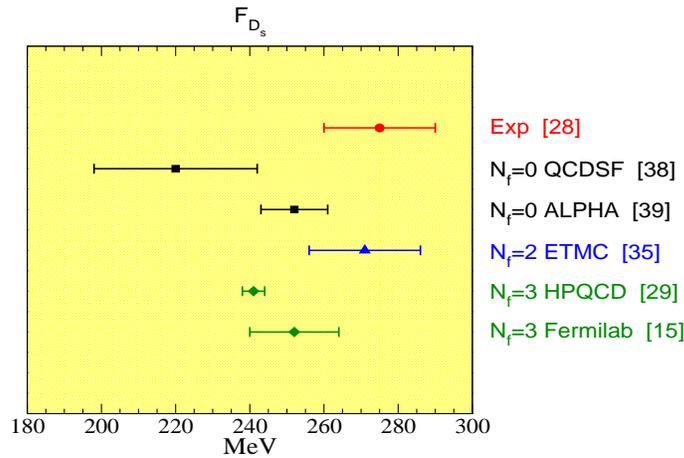}
\caption{Recent determinations of the $D_{\rm s }$ meson decay constant compared to the experimental 
result.} 
\label{fig:FD}
\end{center}

\vspace{-.5cm}
\end{figure}

The different determinations of  the $D_{(\rm s)}$ decay constant show statistically significant quenching effects. 
For $F_{\rm D_{\rm s}}$, which has been computed by  most of the collaborations, 
the results discussed are collected in figure~\ref{fig:FD}. There the errors have been conservatively added linearly.
The figure also shows the tension, which is emerging with the latest experimental measurement 
($F_{\rm D_{\rm s}}=275(10)(5)$ MeV and $F_{\rm D_{\rm s}}/F_{\rm D}=1.24(10)(3)$ ) from CLEO-c~\cite{Stone:2007dy}.
The lattice determinations are  indeed systematically below it and in some cases the discrepancy is above two standard
deviations.
As discussed, for the $B_{(\rm s)}$ system quenching effects appear even larger. This shouldn't be puzzling as 
for B-physics effective theories, rather than relativistic QCD, are simulated.  The inclusion of dynamical fermions
can therefore have different effects in the two cases.
\section{$B_{\rm (s)}-\bar{B}_{\rm (s)}$ mixing}
The weak interactions induce mixings among flavor eigenstates. At low energies and for B-mesons the process is
described by the $\Delta B=2$ Weak Effective Hamiltonian. In particular the matrix elements of four-fermion
operators $O^{\Delta B=2}$ (corresponding to the box diagrams) among meson ($B_{\rm (s)}$ and $\bar{B}_{\rm (s)}$) 
states need to be computed. The mixing is expressed through the oscillation frequency $\Delta m_{(\rm s)}$
\be
m_{\rm B_{\rm q}} \Delta m_{\rm q} \propto |V_{\rm tq}^*V_{\rm tb}|^2 \langle 
\overline{B}_{\rm q} | O_{\rm VV+AA} |B_{\rm q} \rangle \; ,
\ee
where the proportionality factor is given by the Wilson coefficients (functions of $m_{\rm t}/m_{\rm W}$ and
$G_{\rm F}$).
It is customary to introduce the $B_{\rm B_{(\rm s)}}$ parameter by dividing out the result in the vacuum-saturation 
approximation
\be 
{{3}\over{8}} {{\langle \overline{B}_{\rm q} | O_{\rm VV+AA} |B_{\rm q} \rangle}
\over{F_{\rm B_{\rm q}}^2 m_{\rm B_{\rm q}}^2 }} = B_{\rm B_{\rm q}} \; .
\ee
Oscillations of  $B$ mesons are comparatively ``slow'' and have been observed since UA1, the PDG~\cite{PDG} average 
for $\Delta m$ is $0.507(5)$ ps$^{-1}$.
On the contrary $B_{\rm s}-\bar{B}_{\rm s}$ mixing is very fast and $\Delta m_{\rm s}$ has been measured only recently 
by CDF~\cite{Abulencia:2006ze}, with the result $\Delta m_{\rm s}=17.77(10)(7)$ ps$^{-1}$. Notice that the accuracy of
both measurements is at the percent level, which will be very difficult to match from the theoretical side.   
However, by combining these experimental determinations with the lattice computations of the $B_{\rm B_{(\rm s)}}$ 
parameters the Standard Model values for $V_{\rm td}$ and $V_{\rm ts}$ (or ratios thereof) could be extracted.

This year three Collaborations have reported results on the B-parameters with three dynamical flavors.
The HPQCD Collaboration in~\cite{Dalgic:2006gp} has computed $B_{\B_{\rm s}}$ and also the matrix elements
for $\Delta \Gamma_{\rm s}$ on the MILC staggered AsqTad configurations at $a\simeq 0.12$ fm, in a volume 
$20^3 \times 64$ and for 
sea quark masses equal to one half and one quarter of the strange quark mass. The b-quark is treated using NRQCD. 
The results show very little dependence on the light quark masses within the errors and the final estimate is
$F_{\B_{\rm s}} \sqrt{ B_{\B_{\rm s}}^{\rm {RGI}}}=281(21)$ MeV and 
$B_{\B_{\rm s}}(m_{\rm b})=0.76(11)$ using two-loop formulae for the 
conversion to the $\overline{\rm MS}$ scheme. In the computation the operators in QCD are related to their NRQCD
counterparts including O($1/m_{\rm b}$) corrections, which bring in operators of dimension seven. 
These operators require a power divergent subtraction, which  in~\cite{Dalgic:2006gp} is performed at the one-loop
level. This means that the subtracted operator is still power divergent.
With the one-loop value for the coefficient the subtraction itself is about $10\%$ of the final result on 
$B_{\B_{\rm s}}$ and it gives the
largest contribution to the systematical error (see table~2 in~\cite{Dalgic:2006gp}). It is clear that the situation
becomes worse as finer lattice resolutions are considered, as the subtraction grows linearly with $1/a$. A computation
of the subtraction coefficient to higher orders in perturbation theory could at least help in reducing the systematic
uncertainty associated to the matching. However, as pointed out in~\cite{Davies07}, part of this systematic cancels in 
the ratio $\xi={{F_{\rm B_{\rm s}} \sqrt B_{\rm B_{\rm s}}}\over {F_{\rm B} \sqrt B_{\rm B}}}$, which can be used to
extract $ {{|V_{\rm td}|}\over{|V_{\rm ts}|}} $ from ${{\Delta m_{\rm d}} \over {\Delta m_{\rm s}}}$.
This quantity is now being computed by the HPQCD Collaboration which has presented a study using several time sources
with smearing to reduce the statistical and fitting errors~\cite{Davies07}.

\begin{figure}[htb]
\begin{center}
\includegraphics[width=6.2cm,height=9cm,angle=-90]{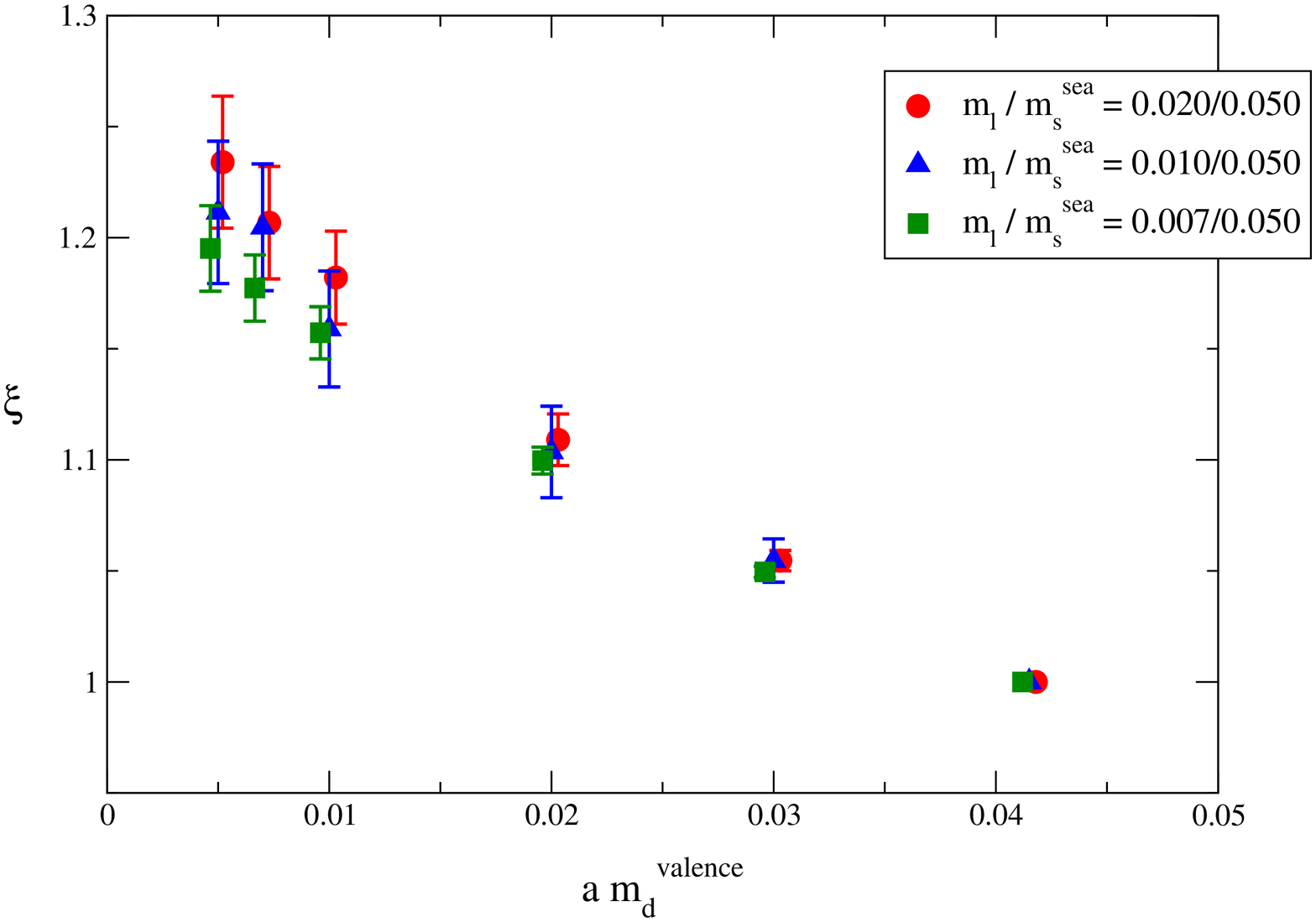}
\caption{$\xi$ as a function of the valence $d$ quark mass and for different values of the sea quark mass
$m_{\rm l}$. Figure from~\cite{Gamiz07}.}
\label{fig:Gamiz}
\end{center}

\vspace{-.5cm}
\end{figure}

The Fermilab-MILC Collaborations reported about the work in progress on the computation of the ratio 
$\xi$~\cite{Gamiz07} employing the Fermilab formalism for heavy quarks and again the  MILC configurations generated
at $a\simeq 0.12$ fm. Matching and renormalization (also including O($1/m_{\rm b}$)) are implemented in one-loop
perturbation theory. The preliminary results are shown in figure~\ref{fig:Gamiz} (statistical errors only).
The light sea quark mass dependence seems rather small compared to the statistical error, whereas the dependence on
the light valence quark mass is noticeable within statistics.
To finalize the results the S$\chi$PT formulae for the relevant hadronic matrix elements are being determined 
in order to be able to simultaneously fit the results for different quark masses and lattice spacings.
Indeed the Collaborations plan to repeat the computation on a finer and a coarser lattice.

The RBC and UKQCD Collaborations have implemented HQET at the leading order (static approximation) combined with
light domain wall fermions for a computation of the mixing parameters with 2+1 dynamical flavors~\cite{Jan07}.
The lattice used has a linear extent $L\simeq 2$ fm with $a\simeq 0.12$ fm and $L_{\rm s}=16$, which  for 
the residual mass from the five-dimensional Ward identity gives 
$am_{\rm res}=0.00308(4)$. Three values of the light sea quark mass have been considered, such that
the lowest pion mass reached is 400 MeV, while for the static quark  the APE and HYP2~\cite{Della Morte:2005yc}
discretizations have been used. 
The preliminary results $F^{\rm stat}_{\B_{\rm s}}=220(32)$ MeV, 
$F^{\rm stat}_{\B_{\rm s}}/F^{\rm stat}_{\B}=1.10(^{+11}_{-5})$, 
$B^{\rm stat}_{\B_{\rm s}}(m_{\rm b})=0.79(4)$ and $B^{\rm stat}_{\B}(m_{\rm b})=0.74(10)$ 
(in the $\overline{\rm MS}$ scheme) have been obtained by using 
one-loop mean-field improved estimates of the matching and renormalization factors~\cite{Lok07} and by linearly
extrapolating the data to the physical point.
Large differences between the APE and the HYP2 results have been observed for example for the quantity
$F^{\rm stat}_{\B_{\rm s}}\sqrt{B^{\rm stat}_{\B_{\rm s}}}$ where the discrepancy between the central values is 30\%.
Notice however that even if a chirally invariant light action is used, non-perturbative effects in the 
renormalization constant of the static-light axial current can be large (see figure~\ref{SF_ax_run}) and in addition
static light correlations functions are not automatically O($a$) improved,  
therefore large O($a$) contributions may still affect the results.

The non-perturbative renormalization programme for the parity-odd 
static-light four-fermion operators in the SF scheme has been completed by the ALPHA Collaboration for the quenched
case and for two dynamical flavors.
In all effective theories  the operator $O_{\rm VV+AA}^{\rm QCD}$ is expanded as
\be
O_{\rm VV+AA}^{\rm QCD}(m_{\rm b})=C_{\rm L}(\mu,m_{\rm b}) O_{\rm VV+AA}^{\rm eff}(\mu)+
C_{\rm S}(\mu,m_{\rm b}) O_{\rm SS+PP}^{\rm eff}(\mu) + O(1/m_{\rm b}) \;, 
\ee
in other words, already at leading order, and in the continuum,  the mixing between the two renormalized operators
$O_{\rm VV+AA}^{\rm eff}(\mu)$ and $ O_{\rm SS+PP}^{\rm eff}(\mu)$ has to be considered.
On top of that the bare lattice operators may mix with operators of the same dimension under renormalization. In 
particular if chiral symmetry is broken by the lattice regularization (like with Wilson fermions) the bare operators 
$O_{\rm VV+AA}^{\rm eff}$, $O_{\rm VV-AA}^{\rm eff}$, $O_{\rm SS+PP}^{\rm eff}$ and $O_{\rm SS-PP}^{\rm eff}$
mix among themselves. In the static approximation, it has been shown in~\cite{DellaMorte:2004wn} 
by using symmetry arguments that all the chirality
breaking mixings can be ruled out if one works with Wilson-tmQCD at maximal twist.\footnote{The transformation 
${\mathcal{P}}_{\pi/2}$ introduced in~\cite{DellaMorte:2004wn} is not completely well defined. The conclusion is 
anyway unaffected as the absence of mixings in a mass independent scheme can be proven by using the transformations
${\mathcal{P}}'_{\pi/2}$ and $Ex_5$ only.} 
The renormalization constants needed, in a mass independent scheme, can then be obtained by renormalizing the 
parity-odd  operators $O_{\rm VA+AV}^{\rm stat}$ and $O_{\rm SP+SP}^{\rm stat}$ in the standard Wilson 
case~\cite{Palombi:2005pa} where indeed  $O_{\rm VA+AV}^{\rm stat}$ and $O_{\rm SP+SP}^{\rm stat}$
do not mix with operators of different chirality.

\begin{figure}[htb]
\begin{center}

\vspace{0.2cm}
\includegraphics[width=13.5cm,height=5.8cm]{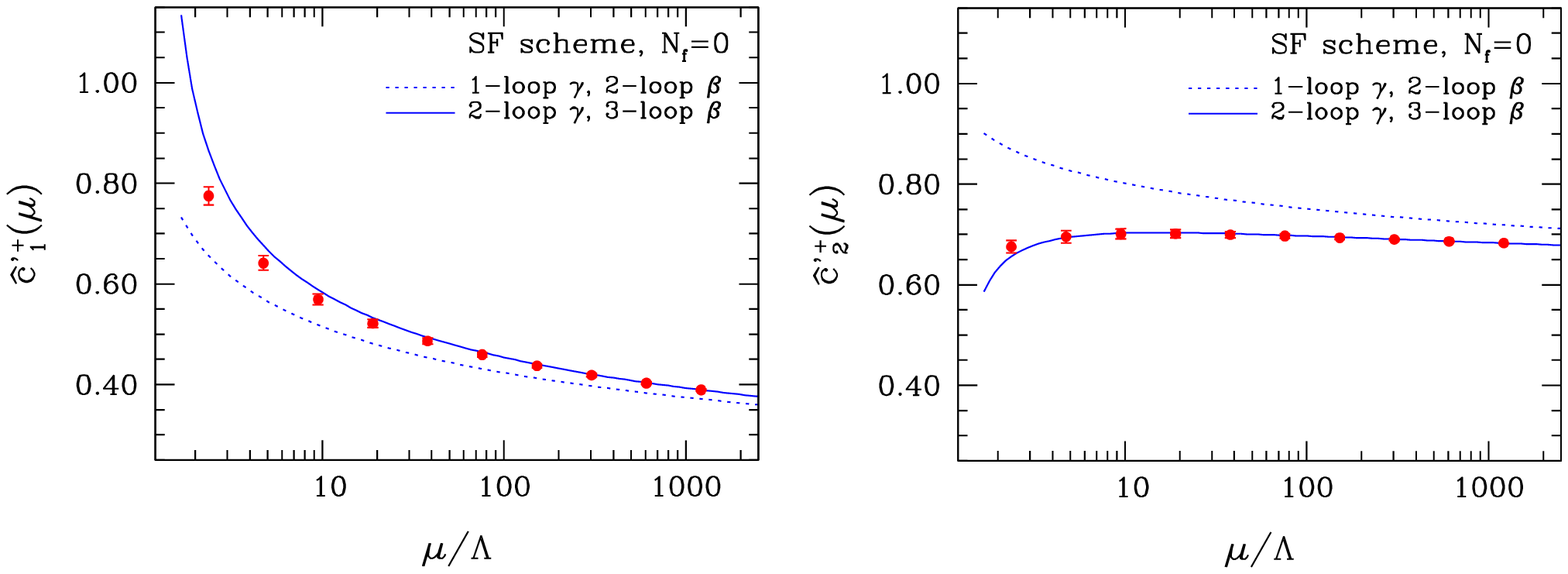}
\caption{Non-perturbative running matrix elements for the operators  $O_{\rm VA+AV}^{\rm stat}$ (left) and
$O_{\rm VA+AV}^{\rm stat}+ 4O_{\rm SP+SP}^{\rm stat}$ (right) in the SF scheme 
(cfr. figure~2).
Figure from~\cite{Palombi:2007dr}.}
\label{fig:c1c2}
\end{center}

\vspace{-.5cm}
\end{figure}

The non-perturbative universal factor relating a matrix element renormalized at the scale $\mu$ in the SF scheme
to the RGI one is shown in figure~\ref{fig:c1c2} for the operators $O_{\rm VA+AV}^{\rm stat}$ and 
$O_{\rm VA+AV}^{\rm stat}+ 4O_{\rm SP+SP}^{\rm stat}$, which renormalize multiplicatively~\cite{Palombi:2007dr}. 
The figure refers to the computation in the quenched theory.
Perturbation theory seems to work for $\mu \geq 1$ GeV for these quantities. In particular for 
$\hat{c}'^+_2$ the series might seem badly convergent from the difference between the 1/2 and the 2/3-loop results, but
quite surprisingly the non-perturbative value eventually agrees with the 2/3-loop one on all the energy range plotted.
Similar findings apply to the $N_{\rm f}=2$ theory as well~\cite{Pena07}. In that case however the final errors on the 
running of the matrix elements are much larger (up to 5\%), which limits somehow the eventual precision one can reach 
on the weak matrix elements. An improvement might be obtained by repeating the calculation at finer lattice spacings.
\section{Form factors for heavy-light and heavy-heavy semi-leptonic decays}
Semi-leptonic decays of $B$ mesons are still the most precise channel for measuring e.g. $|V_{\rm ub}|$. 
On the theoretical side they are described in a well-understood way (compared to hadronic decays) and experimentally
they are easier to study than the less abundant purely leptonic decays. Taking as prototype the $B\to \pi l \nu$
transition, the differential decay rate in the SM reads (ignoring the lepton mass)
\be
{{d\Gamma}\over{dq^2}}={{G^2_{\rm F}}\over{24 \pi^3}} p_{\pi}^3 |V_{\rm ub}|^2 |f_+(q^2)|^2 \;,
\label{ddecay}
\ee
where $q$ is the lepton pair momentum.
The form factor $f_+(q^2)$ can be extracted from the matrix element of the vector current
\be
\langle \pi(p_{\pi}) | V^{\mu} |B(p_{\B}) \rangle = f_+(q^2) (p_{\pi}+p_{\B}+q\Delta_{m^2})^{\mu}+f_0(q^2)q^{\mu} 
\Delta_{m^2}\; ,
\label{fplus_mat}
\ee
with $\Delta_{m^2}=(m_{\B}^2-m_{\pi}^2)/q^2$. For vector to pseudoscalar transitions 
the decay rate is parameterized by four form factors which can be obtained from matrix
elements of the axial and the vector current~\cite{Neubert:1993mb}.

The differential decay rate in eq.~(\ref{ddecay}) grows with the pion momentum and therefore experimental measurements
are more precise for large values of $p_{\pi}$. On the lattice, on the other hand, only the low $p_{\pi}$ 
(or large $q^2$) region is safe from large cutoff effects. Incidentally that is also the region where HQET is applicable.
Notice however that the sensitivity of the matrix element in eq.~(\ref{fplus_mat}) to $f_+(q^2)$ vanishes for 
$q \to q_{\rm max}=(m_{\B}-m_{\pi},\vec{0})$ as the kinematical factor  in front of $f_+(q^2)$ vanishes in that limit.
The form factors are therefore directly computed on the lattice only for some large value of $q^2$  and then 
parameterized over the whole $q^2$ region using functional forms, which include kinematical constraints, HQET scaling 
and dispersion relations as originally proposed in~\cite{Becirevic:1999kt}.

\begin{figure}[htb]
\begin{center}
\vspace{-.5cm}
\includegraphics[width=6.95cm,height=8.4cm,angle=-90]{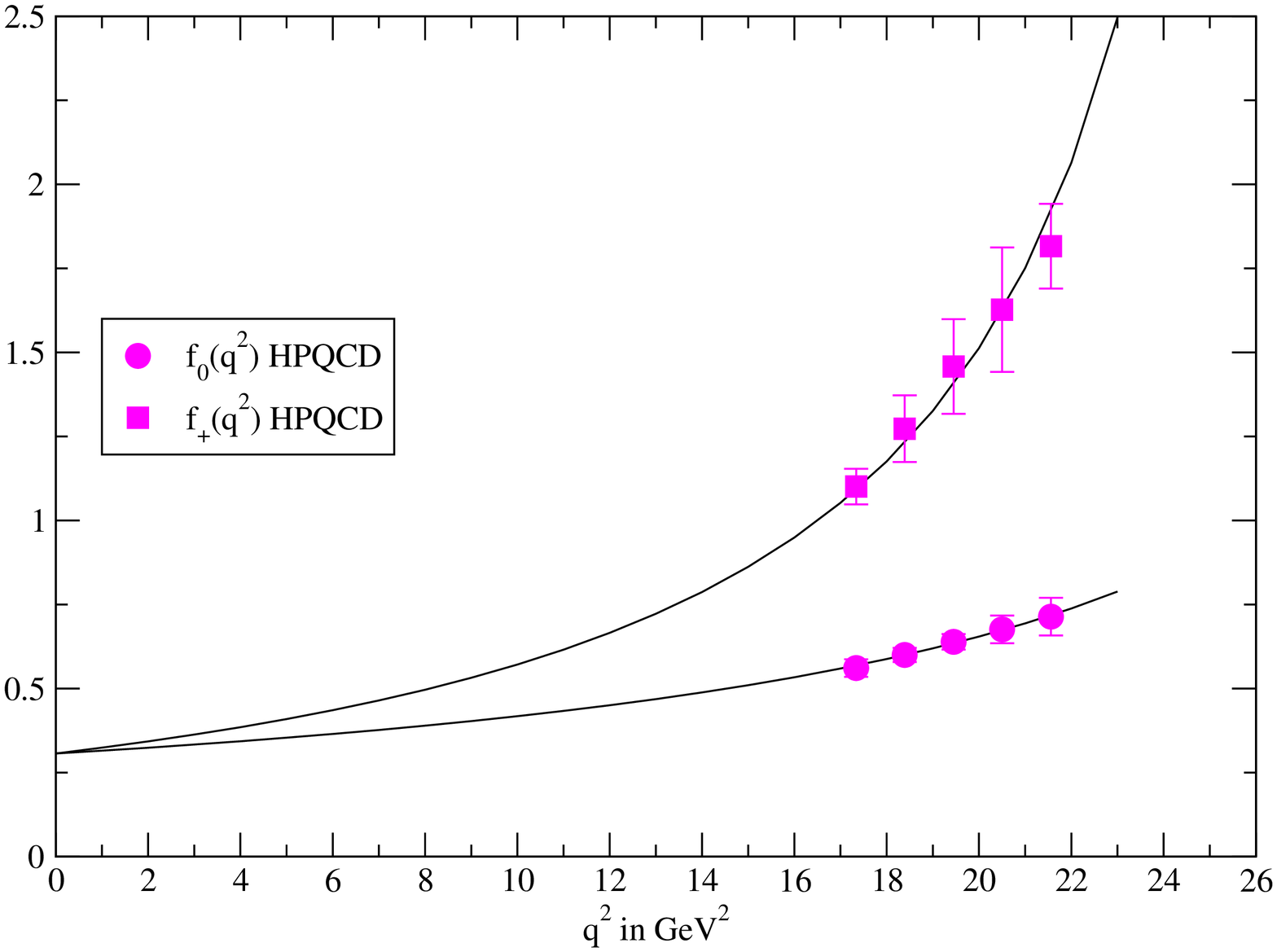}
\caption{Form factors $f_+(q^2)$ and $f_0(q^2)$ from~\cite{Dalgic:2006dt}. Errors are combined statistical and chiral
 extrapolation errors. The curve is the Ball-Zwicky parameterization fit. Courtesy of Junko Shigemitsu.}
\label{fig:formfac}
\end{center}

\vspace{-.5cm}
\end{figure}

The most recent lattice determination of $f_+$ and $f_0$ is due to the HPQCD Collaboration~\cite{Dalgic:2006dt}. 
The three-point correlation functions needed to extract the form factors have been computed 
on the same set of $N_{\rm f}=2+1$ configurations used for measuring $B_{\B _s}$ plus additional sets at lighter
sea quark masses, down to $m_{\rm l}/m_{\rm s} =0.125$. The b-quark has been simulated in the NRQCD formalism with
one-loop matching of the currents to O($1/m_{\rm b}$), i.e. including one-loop subtracted dimension-four operators. 
The subtraction in this case contributes about 5\%  of the final result on the matrix element.
Four lattice momenta have been used for the pion, $\vec{p}_{\pi}=\left\{ (0,0,0), \;(0,0,1), \;(0,1,1)\;\right.$
and  $\left.(1,1,1) \right\} \times {{2\pi}\over{L}}$. For each light quark mass the data are interpolated to fixed 
common values of $E_{\pi}$ and then extrapolated to the physical point using S$\chi$PT and continuum $\chi$PT to assess
the uncertainties in the extrapolation. In the chiral fits the coupling $\gBB$ entering the chiral logs is left free 
to vary in order to use the functional form suggested by  S$\chi$PT also for $E_{\pi}> 2 m_{\pi}$.
The results are plotted in figure~\ref{fig:formfac} together with the curve obtained from the 4-parameter
Ball-Zwicky fit~\cite{Ball:2005tb}.
The errors shown are statistical and chiral extrapolation errors only. As expected from the discussion at the beginning
of this section they grow for large $q^2$ and  statistic is being accumulated to reduce them.
The total error in the final budget is $14\%$, mainly due to statistic, chiral extrapolation and matching of the 
currents. The parameterization of $f_+(q^2)$ is used to obtain
\be
{{1}\over{|V_{\rm ub}|^2}} \int^{q^2_{\rm max}}_{16 {\rm GeV}^2} {{d\Gamma}\over{dq^2}}dq^2=2.07(41)(39)\; 
{\rm ps}^{-1} \;,
\ee
which combined with the experimental result from the HFAG~\cite{Barberio:2006bi} 
for the integrated decay rate in the equation above gives $|V_{\rm ub}|=3.55(25)(50)\times 10^{-3}$.
The tension with the inclusive determination $|V_{\rm ub}|=4.49(33)\times 10^{-3}$, which in the SM 
is dis-favored by the global Unitarity Triangle fits~\cite{Lubicz:2007yu} and poses problems also for Minimal 
Flavor Violating extensions of the SM~\cite{Altmannshofer:2007rj}, is still there.

\begin{figure}[htb]
\vspace{-.6cm}
\begin{center}
\includegraphics[width=6.2cm,height=8cm,angle=-90]{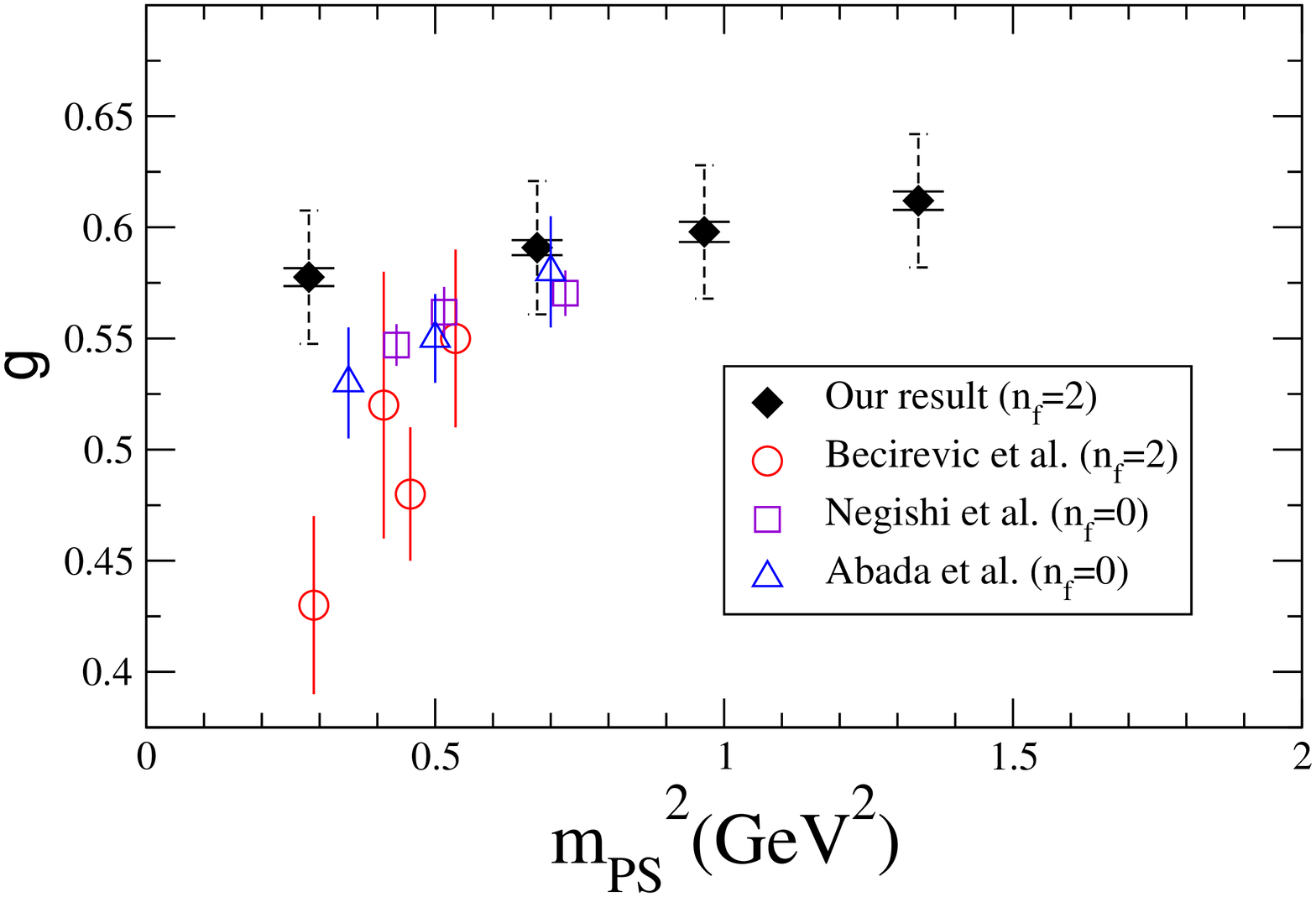}
\vspace{-.25cm}
\caption{Compilation of results for $\hat{g}=g$ from~\cite{Negishi:2006sc,Onogi07,Abada:2003un,Becirevic:2005zu}. 
The two error bars on the diamonds correspond to statistical and perturbative errors. The empty squares and triangles 
are non-perturbatively renormalized. Figure from~\cite{Onogi07}.}
\label{fig:gBB}
\end{center}

\vspace{-.4cm}
\end{figure}

An alternative and complementary approach, which can provide precise results for large $q^2$, consists in using
Heavy Flavor $\chi$PT. At leading order in chiral perturbation theory and through the order $1/{m_{\rm b}}$
in the heavy quark expansion~\cite{Boyd:1994pa}
\be
f_+(q^2)=-{{F_{\B^*}}\over{2F_{\pi}}}\left[\gBB \left({{1}\over{v\cdot
k_{\pi} -m_{\B^*}+m_{\B}}}-{{1}\over{m_{\B}}}\right)+{{F_{\B}}\over{F_{\B^*}}}\right]\;, 
\ee
where $v$ is the velocity of the heavy meson (notice $[F_{\B^*}]=2$ in the formula).
The method requires a computation of the coupling $\gBB$, which in the static approximation is obtained
from the matrix element of the light-light axial current between a $B$ and a $B^*$ state at zero 
momentum (and is called $\hat{g}$).
A very precise determination of $\hat{g}$ in the quenched approximation has appeared this year in~\cite{Negishi:2006sc},
while a preliminary $N_{\rm f}=2$ result has been presented at this conference~\cite{Onogi07}.
In both cases the HYP1 static action~\cite{Della Morte:2005yc} has been used and the required two- and three-point 
correlation functions have been evaluated adopting the all-to-all techniques introduced in~\cite{Foley:2005ac}. 
In the quenched approximation 100 eigenvectors have been
computed for the low-mode part of the correlators whereas in the dynamical case 200 eigenvectors were needed,
the number of configurations used, on the other hand, was only 32 and 100, respectively.
The lattice spacings in both cases were quite coarse, $0.1$ fm in the quenched computation and $0.2$ fm for
$N_{\rm f}=2$. The results are collected in figure~\ref{fig:gBB} (from~\cite{Onogi07}) 
as a function of the pseudoscalar meson mass.
In the $N_{\rm f}=0$ case the final value $\hat{g}=0.517(16)$ is obtained by extrapolating linearly the data in 
$a^2m_{\pi}^2$ while for the preliminary dynamical result $\hat{g}=0.55(1)(3)(3)(6)$ different extrapolations have been 
compared. For the latter value the first error is statistical, the second from the chiral extrapolation, the third 
from the renormalization factor (computed at one-loop only) and the fourth is an estimate of discretization effects.

\begin{figure}[htb]

\begin{center}
\includegraphics[width=8.5cm,height=6cm]{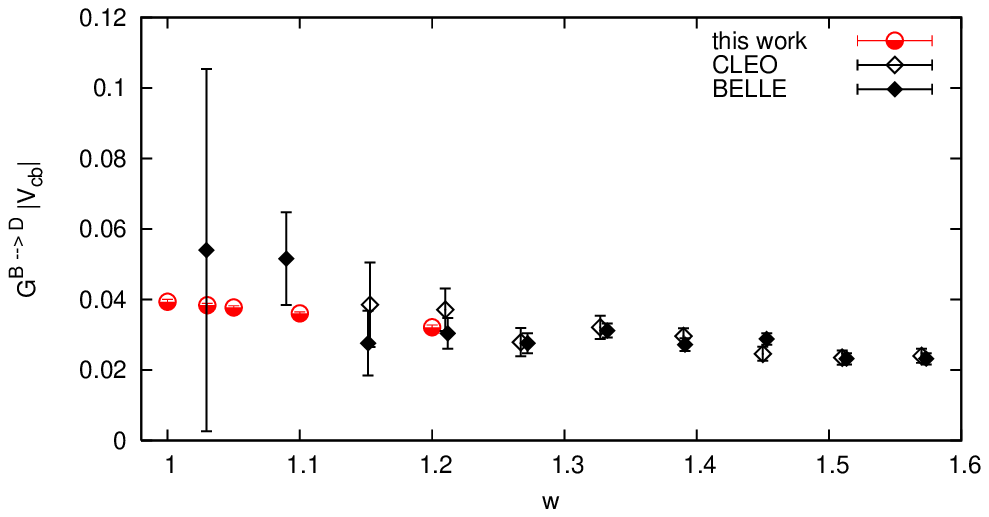}
\caption{Comparison of $|V_{\rm cb}|G(w)$ with experimental data~\cite{Bartelt:1998dq,Abe:2001yf}. 
The form factor has been computed 
in~\cite{de Divitiis:2007ui} and the figure has been obtained by extracting $|V_{\rm cb}|$ at $w=1.2$, which yields 
$|V_{\rm cb}|=3.84(9)_{\rm theo}(42)_{\rm exp}\times 10^{-2}$. Figure from~\cite{de Divitiis:2007ui}.}
\label{fig:BDlnu}
\end{center}

\vspace{-.5cm}
\end{figure}

Considering now heavy to heavy transitions, the Rome~II group in~\cite{de Divitiis:2007ui} presented a quenched
computation of the form factor $G(w)$, where $w$ is the scalar product of the velocities 
of the meson in the initial and in the final state, for the $B \to D l \nu$ decay. 
The computation makes use of the step scaling method developed by the
group in order to avoid resorting to effective theories. In a small volume of linear size $L_0=0.4$ fm and a very
fine lattice resolution the form factor $G(w,L_0)$ is computed at the physical values of the bottom and charm quark 
masses. This number is of course affected by very large finite size effects, which are removed
by multiplying (twice) by the step scaling function 
$\sigma(L,s,w,m_{\rm h})$ defined (for a heavy, would be bottom, quark of mass $m_{\rm h}$) as 
\be
\sigma(L,s,w,m_{\rm h})={{G(sL,w,m_{\rm h})}\over{G(L,w,m_{\rm h})}} \; , \quad s>1 \;.
\ee
For $L>L_0$ the step scaling function can not be computed directly at the physical value of the b-quark mass, and the
key idea of the approach is exactly that it is enough to compute it for 
$m_{\rm h} \simeq {{L_0}\over{L}}m_{\rm b}$ (which  typically means that in the last step $m_{\rm h} \simeq m_{\rm c}$)
and then extrapolate it in $1/m_{\rm h}$ to $m_{\rm b}$. The extrapolation is expected  to be smooth as finite size 
effects (which is what the step scaling function describes) shouldn't depend strongly on the heavy-mass scale.  
This is found to be true in all applications of the method (see~\cite{Guazzini:2007ja} for
recent ones where the step scaling functions have been computed also in HQET to turn the extrapolations into
interpolations). For the case at hand the form factor is finally obtained in a ($1.2$ fm$)^3$ volume as
\be
G(w)=\sigma(2L_0,1.5,w,m_{\rm b}) \sigma(L_0,2,w,m_{\rm b}) G(w,L_0) \;.
\ee
Each factor is computed in the continuum limit (although extrapolating in $a^2$ from two lattice resolutions only 
for the $\sigma$'s) and the product is then linearly extrapolated in the light quark mass from masses above
 $m_{\rm s}/4$. Different values of $w$ have been considered by adopting flavor twisted boundary conditions.
The result is shown in figure~\ref{fig:BDlnu}.

The ETM Collaboration also computed the form factors for heavy pseudoscalar to pseudoscalar 
transitions~\cite{Simula07} in the same $N_f=2$ setup used for the computation of 
$F_{\rm D}$ and $F_{\rm D_{\rm s}}$~\cite{Blossier:2007pt} (i.e. with heavy quarks around the charm).
The preliminary result at $a\simeq 0.09$ fm is consistent with the one in~\cite{de Divitiis:2007ui} within the 
still rather large statistical errors.

We close this section with the computation of the form factor for $B\to D^* l \nu$ at zero recoil from
the Fermilab and MILC Collaborations~\cite{Laiho07}. This channel is less helicity suppressed than
the $B\to D l \nu$ one and it is therefore preferred for the extraction of $|V_{\rm cb}|$.
The computation simplifies in the zero recoil kinematics as in this limit only one, usually called $h_{\rm A_1}$,
of the four form factors contributes. It is obtained from the matrix element of the heavy-heavy axial
current between $B$ and $D^*$ states. 
As proposed in~\cite{Laiho07} the form factor can actually be computed directly from the double ratio
\be
{\mathcal{R}_{\rm A_1}}={{\langle D^* | \bar{c}\gamma_j\gamma_5 b| \overline{B} \rangle
\langle \overline{B} | \bar{b}\gamma_j\gamma_5 c | D^* \rangle}\over
{\langle D^* | \bar{c}\gamma_4 c | D^* \rangle
\langle \overline{B} | \bar{b}\gamma_4 b | \overline{B} \rangle }}=|h_{\rm A_1}|^2 \;.
\label{eq:dr}
\ee
Three similar double ratios had been introduced in~\cite{Hashimoto:2001nb} in order to compute $h_{\rm A_1}(1)$
to O$(1/m_{\rm b}^2)$ in the heavy quark expansion. The expression in eq.~(\ref{eq:dr}) gives the correct
answer to all orders and preserves the feature that most of the lattice current renormalizations
cancel in the ratio. Notice however that contrary to the double ratios
in~\cite{Hashimoto:2001nb},  ${\mathcal{R}_{\rm A_1}}$ has a non-trivial value different from one
already for $m_{\rm b}=m_{\rm c}$ and therefore the uncertainty on it doesn't strictly scale
as ${\mathcal{R}_{\rm A_1}}-1$.

\begin{figure}[htb]

\begin{center}
\vspace{.4cm}
\includegraphics[width=7.4cm,height=5.4cm]{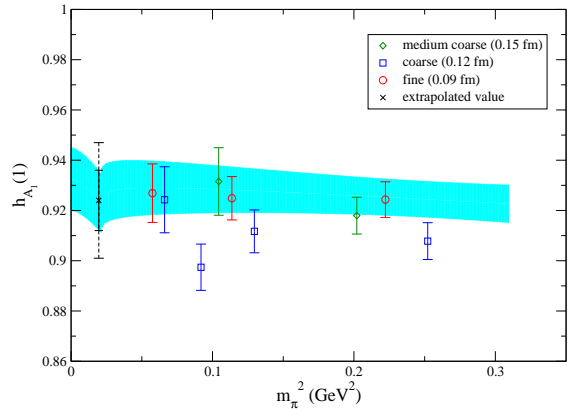}
\caption{Form factor $h_{\rm A_1}(1)$ from~\cite{Laiho07} as a function of the lightest pseudoscalar
meson mass for unitary points. The band is the continuum extrapolated curve and the dashed line on the 
physical point is the total error after the inclusion in quadrature of the systematic ones. Figure taken 
from~\cite{Laiho07}.}
\label{fig:BDslnu}
\end{center}

\vspace{-.5cm}
\end{figure}

In~\cite{Laiho07} the method has been applied on the $N_{\rm f}=2+1$ MILC rooted staggered configurations
together with the Fermilab formalism for heavy quarks. The results are collected in figure~\ref{fig:BDslnu}.
The physical, continuum  value obtained by using S$\chi$PT formulae for the chiral/continuum extrapolations is 
$h_{\rm A_1}(1)=0.924(12)(19)$ where the second error is the sum in quadrature of all the systematic ones.
By combining it with the experimental measurement (see~\cite{Barberio:2006bi}), the estimate
$|V_{\rm cb}|=3.87(9)_{\rm theo}(7)_{\rm exp}\times 10^{-2}$ is obtained.

\section{b-quark mass and B meson decay constant in HQET at O($1/m_{\rm b}$)}
HQET on the lattice was introduced in~\cite{HQET,Eichten:1989zv} twenty years ago.
It offers a theoretically very sound approach to non-perturbative B-physics as it provides
the correct asymptotic description of QCD correlation functions in the limit
$m_{\rm b} \to \infty$. Subleading effects are described by higher dimensional operators
whose coupling constants are formally O$(1/m_{\rm b})$ to the appropriate power.
The theory can be treated in a completely non-perturbative way including renormalization and matching,
in principle to an arbitrary order in $1/m_{\rm b}$, as it was shown in~\cite{Heitger:2003nj}.
This implies the existence of the continuum limit at any fixed order in the expansion.
However precise computations have been hampered for a long time by the poor signal to noise ratio
in heavy-light correlation functions at large time separations, which affects the Eichten-Hill action.
The signal can be exponentially improved by considering minimal modifications of the action, where the link in
the time covariant derivative is replaced by a smeared link~\cite{DellaMorte:2003mn,Della Morte:2005yc}.
The inclusion of dynamical quarks is straightforward and the approach can be used together with other methods, as
for example the one proposed by the Rome II group (see~\cite{Guazzini:2007ja} for such recent applications).

To fix the notation we write the HQET action at O($1/m_{\rm b}$) as
\be
S_{\rm HQET}=a^4 \sum_x \left\{\bar{\psi}_{\rm h}(D_0 +\delta_{\rm  m}
  )\psi_{\rm h}+\omega_{\rm spin} \bar{\psi}_{\rm h}(-\sigma  {\bf B})\psi_{\rm h} + \omega_{\rm kin}
  \bar{\psi}_{\rm h}\left(-{\bf D}^2\right) \psi_{\rm h} \right\}
\label{eq:HQET}
\ee
with $\psi_{\rm h}$ satisfying $P_+ \psi_{\rm h}=\psi_{\rm h}$ and $P_+={{1+\gamma_0} \over {2}}$.
The parameters $\omega_{\rm spin}$ and $\omega_{\rm kin}$ are formally O($1/m_{\rm b}$).
For the computation of the b-quark mass the task is to fix  $\delta_{\rm m}$, $\omega_{\rm kin}$
and $\omega_{\rm spin}$ non-perturbatively by performing a matching to QCD.
Actually, by considering spin averaged quantities we can immediately get rid of the contributions
proportional to $\omega_{\rm spin}$. I will give here a short overview of the computation and present the final
results, precise definitions can be found in the corresponding publications~\cite{Della Morte:2006cb,Garron07}.
Let us start by remarking that in order not to spoil the asymptotic convergence of the series the matching must be done 
non-perturbatively (at least for the leading, static piece) as soon as the $1/m_{\rm b}$ corrections are included.
Following~\cite{Sommer:2006sj}, one can imagine having computed a matching
coefficient $C_{\rm match}$ for the static theory at order $l-1$ in perturbation theory. The truncation error
$\Delta C_{\rm match}$ is
\be
\Delta C_{\rm match} \propto [\bar{g}^2(m_{\rm b})]^l \simeq \left\{ {{1}\over{2b_0 \ln (m_{\rm b}/\Lambda_{\rm QCD})}}
 \right\}^l \gg{{\Lambda_{\rm QCD}}\over{m_{\rm b}}} \;\;\;\;{\rm as}\;\; m_{\rm b}\to \infty\;,  
\ee
where $\bar{g}^2$ is a renormalized coupling at the scale $m_{\rm b}$ and $b_0$ is the first coefficient of the $\beta$
function. In other words the perturbative error due to the matching coefficient of the static term is much larger
than the power corrections in the large $m_{\rm b}$ limit.
In our framework matching and renormalization are performed simultaneously and non-perturbatively.

As the action
in eq.~(\ref{eq:HQET}) would produce a non-renormalizable theory, we treat the $1/m_{\rm b}$ corrections to the static,
renormalizable theory as space-time insertions in correlations functions. For correlation functions of some
multilocal fields $O$ this means
\be
\langle O \rangle =\langle O \rangle_{\rm stat} +\omega_{\rm kin} a^4 \sum_x\langle O O_{\rm kin}(x) \rangle_{\rm stat}
+\omega_{\rm spin} a^4 \sum_x\langle O O_{\rm spin}(x) \rangle_{\rm stat}\; ,
\ee
where $\langle O \rangle_{\rm stat}$ denotes the expectation value in the static approximation and 
$O_{\rm kin}(x)$ and $O_{\rm spin}(x)$ are given by $\bar{\psi}_{\rm h}(x) {\bf \sigma B} \psi_{\rm h}(x)$
and $\bar{\psi}_{\rm h}(x) {\bf D}^2 \psi_{\rm h}(x)$, respectively.
We work with Schr\"odinger functional boundary conditions, i.e. we consider QCD with Dirichlet boundary
conditions in time and periodic boundary conditions in space (up to a phase $\theta$ for the fermions).
For the computation in~\cite{Della Morte:2006cb} we remain in the quenched approximation.
In a small volume of extent $L_1 \simeq 0.4$ fm, one can afford lattice spacings $a$ sufficiently smaller than 
$1/m_{\rm b}$, in such a way that the b-quark propagates correctly up to discretization errors of O($a^2$).
QCD observables defined in this volume are described in HQET up to effects of
O$\left({{\Lambda_{\rm QCD}}\over{m_{\rm b}}}\right)^2$ and O$\left({{1}\over{L_1m_{\rm b}}}\right)^2$. The size 
$L_1$ is chosen in order to have the two effects of the same size.
We consider two quantities, $\Phi_1^{\rm QCD}(L,m_{\rm h})$ defined exploiting the sensitivity of SF-correlation 
functions to the angle $\theta$ and  $\Phi_2^{\rm QCD}(L,m_{\rm h})$, which is given by 
$L\Gamma_1$ where $\Gamma_1$ is a finite volume effective energy. When expanded in HQET\footnote{We set
the mass counterterm $\delta_{\rm m}$ in the action to zero here. Its contribution is taken into account in the overall
energy shift $m_{\rm bare}$ between the effective theory and QCD.}, $\Phi_1^{\rm HQET}(L)$ is given 
by $\omega_{\rm kin}$ times a quantity defined in the effective theory (which we call $R_1^{\rm kin}(\theta,\theta')$)
while $\Phi_2^{\rm HQET}(L)$ is a function of $\omega_{\rm kin}$ and $m_{\rm bare}=
\delta_{\rm m}+m_{\rm h}$ involving two other HQET quantities ($\Gamma_1^{\rm stat}$ and $\Gamma_1^{\rm kin}$) . 
Obviously, by equating  $\Phi_{\rm i}^{\rm QCD}(L_1,m_{\rm h})$ to $\Phi_{\rm i}^{\rm HQET}(L_1)$ one can determine
the bare parameters $m_{\rm bare}$ and $\omega_{\rm kin}$ as functions of $m_{\rm h}$ at the lattice spacings
used for the volume $L_1^3$. To eventually compute $m_{\rm b}$ we need the phenomenological,
large volume, input of the spin-averaged vector-pseudoscalar B-meson mass, $m_{\rm B}^{\rm av}$.
Here we introduce the step scaling functions $\sigma_{\rm ij}(L)$ to evolve the $\Phi_{\rm i}$'s
to larger volumes and write
\be
\Phi_{\rm i}^{\rm HQET}(2L_1, m_{\rm h})=\sum_{\rm j}\sigma_{\rm ij}(L_1)\Phi_{\rm j}^{\rm QCD}(L_1,m_{\rm h}) +
\delta_{{\rm i}2}\sigma_{\rm m}(L_1) \;.
\ee
Notice that $\Phi_{\rm i}^{\rm HQET}(2L_1, m_{\rm h})$ constructed in this way still has a dependence on $m_{\rm h}$, 
which is inherited from the matching to QCD in $L_1$.
The step scaling functions on the other hand are defined in HQET and have a continuum limit there.
After two evolution steps volumes of extent roughly $1.5$ fm are reached and the bare parameters $m_{\rm bare}$
and $\omega_{\rm kin}$ can be computed again {\it as functions of $m_{\rm h}$} for the corresponding lattice spacings.
They are expressed in terms of step scaling functions, $\Phi_{\rm i}^{\rm QCD}(L_1,m_{\rm h})$ and  
quantities computed in HQET (the large volume version of $R_1^{\rm kin}(\theta,\theta')$, $\Gamma_1^{\rm stat}$ and 
$\Gamma_1^{\rm kin}$). At this point the b-quark mass can finally be determined by solving for $m_{\rm h}$
the equation
\be
m_{\rm B}^{\rm av}= E^{\rm stat} + \omega_{\rm kin}(m_{\rm h}) E^{\rm kin} + m_{\rm bare}(m_{\rm h}) \; ,
\label{eq:largeV}
\ee
where  $E^{\rm stat} =\lim_{L \to \infty} \Gamma_1^{\rm stat}$ and
$E^{\rm kin}= -\langle B| a^3 \sum_{\bf z} O_{\rm kin}(0,{\bf z})| B \rangle_{\rm stat} $ with $\langle B|B \rangle=1$.
In eq.~(\ref{eq:largeV}) I have emphasized the dependence of the bare parameters $m_{\rm bare}$ and $\omega_{\rm kin}$
on $m_{\rm h}$.  However when those are re-expressed in terms of $\Phi_{\rm i}^{\rm QCD}(L_1)$, $\sigma_{\rm ij}$,
$R_1^{\rm kin}(\theta,\theta',L_2)$, $\Gamma_1^{\rm stat}(L_2)$ and $\Gamma_1^{\rm kin}(L_2)$  the equation involves 
only quantities which have a continuum limit either in QCD or HQET. This in particular implies that in the procedure 
all power divergences have been non-perturbatively subtracted.

To see how the different pieces combine together in the final result it is instructive to consider more explicitly
the relatively simple case of the computation in the static approximation. In this situation only the parameter
$m_{\rm bare}$ needs to be determined. In the small volume the matching condition reads
\be
\Gamma_1(L_1,m_{\rm h})=\Gamma_1^{\rm stat}(L_1)+m_{\rm bare} \;,
\label{eq:1}
\ee
and its large volume version is
\be
m_{\rm B}^{\rm av}=E^{\rm stat} + m_{\rm bare} \;,
\label{eq:2}
\ee
to this order we could have just as well used $m_{\rm B}$ or $m_{\rm B^*}$ in the previous equation.
If we were able to simulate the small and the large volumes at the same lattice spacings we could insert
$m_{\rm bare}$ from eq.~(\ref{eq:1}) into eq.~(\ref{eq:2}) and obtain the master equation
\be
m_{\rm B}^{\rm av}=(E^{\rm stat}-\Gamma_1^{\rm stat}(L_1))+\Gamma_1(L_1,m_{\rm h}) \;,
\ee
whose solution is the b-quark mass in the static limit. To circumvent the problem we bridge the gap in volume
in two steps by inserting a step scaling function $\sigma_{\rm m}(L_1)=2L_1(\Gamma_1^{\rm stat}(2L_1)-
\Gamma_1^{\rm stat}(L_1))$ into the master equation (i.e. we add and subtract $\Gamma_1^{\rm stat}(2L_1)$), 
which then becomes
\be
2L_1 m_{\rm B}^{\rm av}-2L_1[E^{\rm stat}-\Gamma_1^{\rm stat}]-\sigma_{\rm m}(L_1)=2L_1\Gamma_1(L_1,m_{\rm h})\;.
\label{eq:master}
\ee
Now the quantities in the square brackets can be computed at the same values of $a$ and their difference has a 
well-defined continuum limit in HQET because in the combination we are non-perturbatively removing all the divergences, 
particularly the linear one.
As announced any reference to bare parameters has disappeared in the final equation.

\begin{figure}[htb!]

\begin{center}
\includegraphics[width=8.7cm,height=6.5cm]{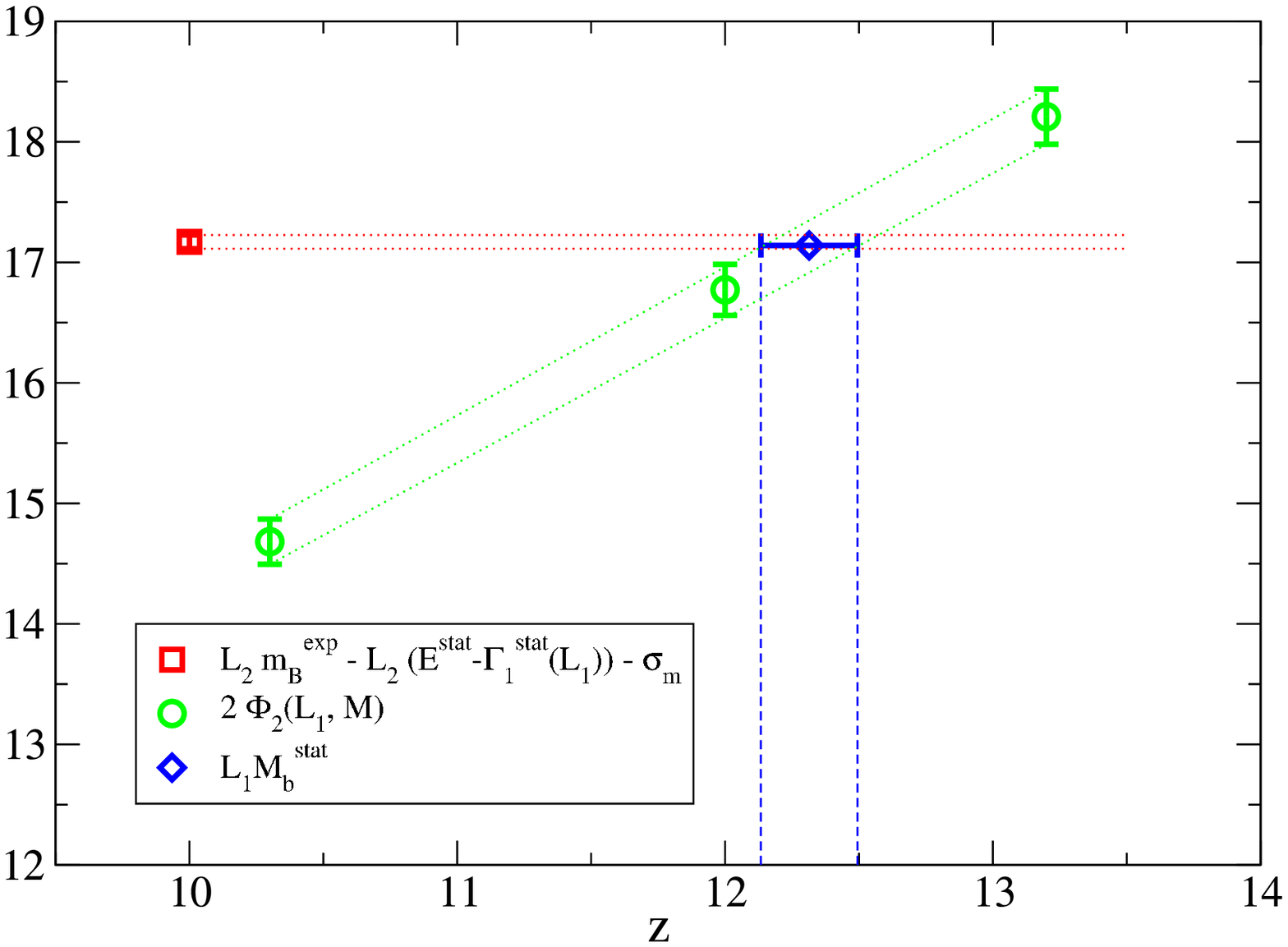}
\caption{Graphical solution of eq.~(5.9) in the quenched approximation. 
In the caption $\Phi_2(L_1,M)=L_1\Gamma_1(L_1,M_{\rm h})$. Data from~\cite{Della Morte:2006cb}.}
\label{fig:master}
\end{center}

\vspace{-.5cm}
\end{figure}

The graphical solution of eq.~(\ref{eq:master}) is shown in figure~\ref{fig:master}.
On the horizontal axis we plot $z=L_1 M_{\rm h}$ where $M_{\rm h}$ is the heavy quark mass in the RGI scheme.
The result is $M_{\rm b}^{\rm stat}=6.806(79)$ GeV using $r_0=0.5$ fm to set the 
scale~\cite{Sommer:1993ce,Guagnelli:1998ud}.

The inclusion of the subleading $1/m_{\rm b}$ effects is more involved and I report here the final results summarized
in table~2
. The different numbers correspond to various matching conditions, identified by the choice 
of the angle(s) $\theta$ and by the strategy adopted, ``main strategy'' for the first line and ``alternative
strategy'' for the second to fourth line. The details can be found in~\cite{Della Morte:2006cb}, what  should be 
emphasized here is that while there are some differences among the static results depending on the matching
condition chosen, those are completely gone once the $1/m_{\rm b}$ terms are included, signalling practically
negligible higher order corrections.
\begin{table}[h!]
\begin{center}\begin{tabular}{cccccc}
\hline
$\theta_0$ &  $r_0\,\Mb^{(0)}$ && \multicolumn{3}{c}{$r_0\,\Mb=r_0\,(\Mb^{(0)} + \Mb^{(1a)} + \Mb^{(1b)})$} \\[1ex]
\hline
& &&
$\theta_1=0$   &  $\theta_1=1/2$ & $\theta_1=1$ \\
& &&
$\theta_2=1/2$ &  $\theta_2=1$   & $\theta_2=0$ \\
\hline
0   & 17.25(20) && 17.12(22)  & 17.12(22)  & 17.12(22) \\
\hline
0   & 17.05(25) && 17.25(28)  & 17.23(27)  & 17.24(27) \\
1/2 & 17.01(22) && 17.23(28)  & 17.21(27)  & 17.22(28) \\
1   & 16.78(28) && 17.17(32)  & 17.14(30)  & 17.15(30) \\
\hline
\end{tabular}
\caption{Results for the RGI mass $M_{\rm b}$ from~\cite{Della Morte:2006cb}.}
\end{center}
\label{tab:mb}
\end{table}
The value eventually quoted in~\cite{Della Morte:2006cb} is $m_{\rm b}(m_{\rm b})=4.347(48)$ MeV in the 
$\overline{\rm MS}$ scheme.

As a further application the decay constant of the $B_{\rm s}$ meson has been computed in quenched QCD including
O$(1/m_{\rm b})$ in HQET~\cite{Garron07}. Four quantities are needed for the matching, which again has been performed
in several different ways. The preliminary results in table~3
show the same pattern discussed for the
b-quark mass. 
\begin{table}[h!]
       \begin{center}
         \begin{tabular}{ccccc}
           \hline
           $\theta_0$ & $F_{{\rm B}_s}^{\rm stat}$ [MeV] &  \multicolumn{3}{c}{
                        $F_{{\rm B}_s}^{\rm stat} + F_{{\rm B}_s}^{\rm (1)}$[MeV]}  \\
           \hline
                      &  & $\theta_1 = 0$ & $\theta_1 = 0.5$ & $\theta_1 = 1$\\
                  &                      &  $\theta_2 = 0.5$ & $\theta_2 = 1$ & $\theta_2 = 0$ \\
           \hline
            $0$   &  $224 \pm 3$  &  $185 \pm 21$  &  $186 \pm 22$  &  $189 \pm 22$\\
            $0.5$ &  $220 \pm 3$  &  $185 \pm 21$  &  $187 \pm 22$  &  $189 \pm 22$\\
            $1$   &  $209 \pm 3$  &  $184 \pm 21$  &  $185 \pm 21$  &  $188 \pm 22$\\
           \hline
         \end{tabular}
         \caption{Results for $F_{\rm B_{\rm s}}$ from~\cite{Garron07}.}
       \end{center}
       \label{tab:FB}
\end{table}
Notice however that the difference at O$(1/m_{\rm b})$  is more significant than the errors suggest as
most of the uncertainties from the large volume part of the computation cancel in the difference. This indeed yields
for instance
\be
F_{\rm B_{\rm s}}^{\rm stat +(1)}(\theta_0=0 ,\theta_1=1 ,\theta_2=0 )-
F_{\rm B_{\rm s}}^{\rm stat +(1)}(\theta_0=1 ,\theta_1=0 ,\theta_2=0.5 )=4\pm 2 \;\; {\rm MeV}\;.
\ee
Finally, the results are in good agreement with the determinations in~\cite{fBstat_Nf0,Guazzini:2007ja}, 
which also go beyond the static approximation.

\section{Conclusions}
A big effort has been devoted in the last years to removing the quenched approximation from lattice computations.
This is absolutely necessary in order to provide precise theoretical estimates to test the Standard Model or to
look for signals of New Physics. 
Several lessons have been learnt from these works. Quenching effects have been proven to be large and chiral
extrapolations to be more delicate than in the $N_{\rm f}=0$ approximation, as partly 
expected~\cite{Becirevic:2006me,Becirevic:2007dg}.

However B-flavor physics is going to become high-precision physics and other 
systematics may significantly affect the results.
I have shown that the uncertainties associated to renormalization, matching, chiral and continuum extrapolations
can easily reach the 5 to 10 percent level. When choosing an approach for performing a first-principle computation the 
possibility to keep these systematics under control should be included among the requirements.

I have described how these problems can be solved non-perturbatively in Heavy Quark Effective Theory on the lattice.
The computations I discussed in this framework on the other hand have been performed in the quenched approximation only 
and therefore the results in principle are not immediately applicable to phenomenology. The extension to dynamical
light fermions is ongoing and first steps have been reported at this conference~\cite{DellaMorte:2007qw}.
The method can be used for several quantities, the b-quark mass and the B-meson decay constant discussed here 
but also $B-\bar{B}$ mixing parameters and form factors for semi-leptonic decays.

More generally, the non-perturbative matching procedure between HQET and QCD in small volume can be adopted also for 
other effective theories, as it has been done in~\cite{Christ:2006us,Lin:2006ur} for a version of the Fermilab action.

\vspace{0.25cm}
{\bf Acknowledgements:} I am grateful to Leonardo Giusti and Rainer Sommer for illuminating 
discussions and for a careful reading of the manuscript. I wish to thank Arifa Ali Khan, 
Benoit Blossier, Nicolas Garron, Hiroshi Ohki, Jack Laiho, Junko Shigemitsu, James Simone,
Silvano Simula and Nazario Tantalo for providing useful material before the conference.

\end{document}